Luminosity

$10^3$
$10^2$
$10^1$
$10^0$
$10^{-1}$

$-0.8$

N(<L)

$10^0$ $10^{-1}$ $10^{-2}$

$F_1$ $F_2$ $R_{core}$

$L_2$

$L_1$

$10^1$
$10^{-1}$
$10^{-3}$
$10^{-5}$

$-0.8$
$-1.5$

N(>F)

$10^0$ $10^1$ $10^2$ $10^3$ $10^4$ $10^5$
Flux

fraction

bright
faint

$T_{50}$ (if > 3 s)

# On the nature of gamma-ray burst time dilations

Ralph A.M.J. Wijers[1] and Bohdan Paczyński[2]
Princeton University Observatory, Peyton Hall, Princeton, NJ 08544–1001, USA

## ABSTRACT

The recent discovery that faint gamma-ray bursts are stretched in time relative to bright ones has been interpreted as support for cosmological distances: faint bursts have their durations redshifted relative to bright ones. It was pointed out, however, that the relative time stretching can also be produced by an intrinsic correlation between duration and luminosity of gamma-ray bursts in a nearby, bounded distribution. While both models can explain the average amount of time stretching, we find a generic difference between them in the way the duration distribution of faint bursts deviates from that of bright ones. This allows us to distinguish between these two broad classes of model on the basis of the duration distributions of gamma-ray bursts, leading perhaps to an unambiguous determination of the distance scale of gamma-ray bursts. We apply our proposed test to the second BATSE catalog and conclude, with some caution, that the data favor a cosmological interpretation of the time dilation.

*Subject headings:* methods: data analysis, statistical — gamma rays: bursts

## 1. Introduction

From the data obtained by BATSE on faint gamma-ray bursts, combined with data from earlier satellites (notably PVO), we know the characteristics of the radial and angular distribution of gamma-ray bursts quite well: we are at or very near the center of the distribution of the known gamma-ray bursts, and the bursts occur at a uniform rate density out to some characteristic distance, after which they become rarer. The big unknown is this characteristic distance, and the two most popular distance models place it in the very outer regions of our Galaxy (Galactic corona models) or at a fair fraction of the distance to the horizon of our universe (cosmological models)[3].

---

[1] E-mail: rw@astro.princeton.edu

[2] E-mail: bp@astro.princeton.edu

The recently announced time dilation of a factor 2 in faint BATSE bursts (Norris et al. 1994), has been adduced as evidence for the cosmological distance scale because the required redshift for the time delay is consistent with the redshift deduced from simple no-evolution, standard-candle models for cosmological gamma-ray bursts (Fenimore et al. 1993, Mao & Paczyński 1992, Piran 1992, Dermer 1992, Wickramasinghe et al. 1993). However, it was quickly pointed out that the effect can also be caused by an intrinsic anti-correlation between luminosity and duration of gamma-ray bursts. Such a correlation can be caused by many things, and relativistic beaming in particular has been discussed in some detail (Brainerd 1994, Mao & Yi 1994, Yi & Mao 1994). Another obvious candidate for introducing such a correlation is requiring a roughly constant total energy output, $E \sim L \Delta t$. Any such anti-correlation will make it possible to get a lengthening of the average burst duration with decreasing flux even in Galactic corona models.

In this paper, we present a way to easily visualize the interplay of distance and luminosity distributions in determining the observed $\log N(>F) - \log F$ of gamma-ray bursts (Sect. 2.). We then use this to show that there is a generic difference between the two above models in the way the duration distribution of faint bursts compares with that of bright ones. We then apply a test based on this difference to the burst summary data as given in the second BATSE catalog, and find that a cosmological interpretation agrees best with the data (Sect. 3.). We conclude by briefly discussing some possible complications (Sect. 4.).

## 2. Interpreting the flux distribution

In Fig. 1 we illustrate how the flux distribution of gamma-ray bursts is shaped by their density and their luminosity function in the case of a Galactic corona model (or any isotropic bounded model in which bursts are nearby enough to make effects of non-Euclidean space negligible; we shall henceforth call all these models 'local'). The central panel shows the luminosity of gamma-ray bursts as a function of their flux. We assume that the luminosity function is independent of distance and, for definiteness, that it is a power law $L^{-\beta}$ extending from $L_1$ to $L_2$, so all bursts in the central panel are between the horizontal dashed lines. The diagonal lines are lines of constant distance, with distance increasing from bottom right to top left. Again for definiteness, we take a density distribution of the form $n(r) \propto (1 + (r/R_{\rm core})^\alpha)^{-1}$.

---

[3]In principle, there is a third, 'Ptolemean' possibility: all gamma-ray bursts can be at the same substantial distance in a shell, and the shape of their flux distribution is simply a reflection of the true luminosity function, which by coincidence has a slope of $-3/2$ at the bright end. All other observed properties and the correlations between them are likewise intrinsic in this case. Since such models are not falsifiable until direct distance measurements to individual bursts become possible, we shall not discuss them further.

There are now two possibilities for making the flux distribution (bottom panel of Fig. 1) agree with the observed $\log N(> F) - \log F$ (a detailed discussion is presented in Ulmer & Wijers 1994). The first is that gamma-ray bursts are effectively standard candles (i.e. if $L_1 \simeq L_2$ or if $\beta > 2.5$ or $\beta < 1$). Then the flattening of $\log N(> F) - \log F$ for weak bursts simply reflects the decreasing density beyond $R_{\rm core}$. The range of luminosities visible at each flux is narrow. The second possibility is that $L_1 \ll L_2$ and $1 < \beta < 2.5$, in which case the slope below the break can simply be the slope of the (cumulative) luminosity function. The slope changes again to a value determined by the density at $F_1$. In this case, the range of luminosities seen at each flux level is quite broad.

## 3. A generic difference in the duration distributions

The durations of gamma-ray bursts range over decades, both at high and low fluxes, so the factor 2 duration difference between faint and bright bursts is a relatively small effect. We therefore assume that most of the duration range of bursts is intrinsic, and seek to explain a non-intrinsic effect on top of that range.

### 3.1. Duration distributions in local models

In this case, the longer average duration of faint bursts is explained by assuming that more luminous bursts tend to have shorter durations. As long as we are on the $-3/2$ part of the flux distribution, the distribution of luminosity (and therefore duration) is independent of flux (Fig. 1). Thus, the duration distribution is the same for all fluxes above $F_2$. But as soon as we go below $F_2$, the effect of decreasing density will remove luminous (i.e. short) bursts from the sample, while leaving the weak (long) bursts unaffected. This effect progresses with decreasing flux until we pass $F_1$. Below $F_1$ all bursts we see are beyond $R_{\rm core}$; if the density distribution is an exact power law, we have once again reached a regime where all fluxes are equivalent, so the duration distribution does not change any further with decreasing flux (see Fig. 1). In case the low-flux slope is set by geometry because the luminosity function is narrow, this would still be true. But now $F_1$ is effectively only just below $F_2$, implying that the duration distribution changes its character over a narrow range in flux just below the break in the slope of $\log N(>F) - \log F$ and remains constant thereafter. (Note that this offers a method, within the limited context of this model, to distinguish narrow luminosity functions from wide ones. Also bear in mind that the change must be substantial even if it occurs over a narrow range in flux in order that the factor of 2 increase in average duration be reproduced.)

The implication is that for non-cosmological interpretations of the observed time dilation, one quite generically expects a change in the average duration of bursts to be the result of removing short bursts from the sample. That is to say, the long end of the gamma-ray burst duration distribution remains unchanged with decreasing flux, but the short end moves up, making the distribution significantly narrower at low fluxes to increase its mean.

### 3.2. Duration distributions in cosmological models

In the case of cosmological distributions there is no need for wide luminosity functions or intrinsic correlations. Gamma-ray bursts can be near-standard candles occurring at a constant rate per unit comoving volume and time (Mao & Paczyński 1992), and their other properties are independent of distance as well. In this case, the burst durations change simply due to redshift stretching. This is clearly different from the duration change discussed above because now the durations of all bursts increase, so the entire distribution shifts to longer durations and the width of the distribution is also increased by $1+z$. This means that the distribution becomes wider with decreasing flux, and its long end shifts up just as well as the short end. The relation between average duration and flux is fixed by the cosmological model one adopts.

### 3.3. Comparing the distributions in practice

In order to test our idea we used the second BATSE catalog (BATSE Team 1994), which is publicly available. Since it only contains summary data, we need to carefully select a sample that is less subject to known biases. This means we want to use peak fluxes in short time bins and use only bursts with durations much longer than those bins in order to avoid incompleteness at short durations and effects of fluence triggering. We therefore used the peak fluxes of bursts on the 64 ms time scale as a measure of brightness, and excluded all bursts shorter than 3 s from our sample. Also, we used the duration measure $T_{50}$ rather than $T_{90}$ (Fishman et al. 1994), because the latter uses the far wings of the burst, which are subject to large errors especially in weak bursts (and are likely to lead to systematic effects in the relative duration measurements of faint and bright bursts). These issues can be dealt with using the full data (as was done by Norris et al. 1994), but not using the catalog. This leaves us with 187 bursts for which all necessary information is available. Following Norris et al., we compare the durations of the brightest bursts (top 10%) with those of a faint group, for which we choose those with count rates ranking from 10% to 20% from the bottom (the faintest, low signal-to-noise group is avoided). The resulting histograms with 19 bursts in each group are shown in Fig. 2. The time dilation between the two groups is evident: the difference in mean $\log(T_{50})$ is $0.34 \pm 0.11$, significant at the $3\sigma$ level and consistent with the factor of 2 dilation found by Norris et al.. However, this was already known, and the issue here is which kind of time dilation is preferred. To address this, we note that the standard deviations of the distributions of $\log(T_{50})$ are $0.316 \pm 0.064$ for the bright sample and $0.356 \pm 0.069$ for the faint one, i.e. the widths are the same. A Kolmogorov-Smirnov test between the two distributions after shifting all faint bursts down in $\log(T_{50})$ by an amount equal to the difference in the sample means yields a probability of 0.79 that the two samples are identical. The unchanged width and the high K-S probability of equality indicate that the duration difference between faint and bright bursts is achieved by an overall shift of the distributions without a change in shape, rather than by a removal of short bursts and a narrowing of the distribution. We therefore conclude that the duration data do indicate that faint bursts are redshifted and that the cosmological distance scale is preferred.

It should be noted again that since the shifts in both models are small relative to the overall width of the duration distribution, it is quite important to account properly for selection effects causing incompleteness of the duration distribution, such as the strong bias against triggering very short bursts. The test is best done on a sample which has a clear maximum in its duration distribution, with the decrease on both sides being real rather than caused by selection effects. We believe that the sample we used satisfies these requirements, even though the choices we made to define our sample are somewhat arbitrary. Small modifications of the sample do not change the result: if we truncate the distributions at 1 s, the difference in $\log(T_{50})$ becomes $0.27 \pm 0.14$. The use of peak count rates in 256 ms bins gives the same difference ($0.35 \pm 0.13$) as the 64 ms bins, but use of 1024 ms bins or $T_{90}$ strongly reduces the difference (typically to $0.12 \pm 0.12$), as expected for reasons mentioned above.

It may be helpful to adopt less arbitrary sample definitions, perhaps by selecting on other burst properties. An example would be to exploit the bimodality discovered in gamma-ray burst properties (Dezalay et al. 1992, Klebesadel 1992, Kouveliotou et al. 1993) and select only the soft group, of which the duration distribution clearly peaks within the range of durations that is well-observed by BATSE. The test can no doubt also be improved by using the actual BATSE time profiles rather than just the summary data from the catalog (see Norris et al. 1994 for a thorough discussion).

## 4. Conclusion and discussion

We have presented a method to distinguish the simplest cosmological and local interpretations of the observed time dilation of faint gamma-ray bursts relative to bright ones. It is based on the fact that the distribution of durations changes with flux in very different ways between the two cases. For cosmological bursts, the entire duration distribution is shifted to longer durations for weak bursts, and the distribution becomes wider by a factor $1 + z$ (hence the distribution of $\log(T)$ has a constant width). For local bursts, the average duration increases by removing short bursts from the sample while leaving the long ones unaffected. This means that the distribution becomes narrower and its long-duration end does not shift up in duration.

The proposed test was applied to a subsample of the second BATSE catalog, and we tentatively conclude that the observed time dilation is cosmological in origin. The effect is quite significant in our test, and we tried to eliminate selection effects and incompleteness as best we could using the summary data in the second BATSE catalog. Given the importance of the issue, we nonetheless caution again that the test can be greatly improved by using the complete burst time profiles in the manner of Norris et al. (1994) to better define samples and test selection effects. Also, since many more bursts have now been found by BATSE than are included in the second catalog, our test can be applied to an independent or larger data set to check that our result is not spurious.

This work was supported in part by NASA Grant NAG 5–1901. RW is supported by a Compton Fellowship (grant GRO/PFP–91–26). This research has made use of data obtained through the Compton Gamma Ray Observatory Science Support Center Online Service, provided by the NASA-Goddard Space Flight Center.

# REFERENCES

BATSE Team 1994, second electronic BATSE catalog, GRO science support center

Brainerd, J. J. 1994, ApJ, in press

Dermer, C. D. 1992, Phys. Rev. Lett., 68, 1799

Dezalay, J., Barat, C., Talon, R., Sunyaev, R., Terekhov, O., & Kuznetsov, A. 1992, in Gamma-Ray Bursts, ed. W. S. Paciesas & G. J. Fishman, AIP Conference Proceedings 265, (American Institute of Physics:New York), 304

Fenimore, E. E., Epstein, R. I., Ho, C., Klebesadel, R. W., Lacey, C., Laros, J. G., Meier, M., Strohmayer, T., Pendleton, G., Fishman, G., Kouveliotou, C., & Meegan, C. 1993, Nature, 366, 40

Fishman, G. J., Meegan, C. A., Wilson, R. B., Brock, M. N., Horack, J. M., Kouveliotou, C., Howard, S., Paciesas, W. S., Briggs, M. S., Pendleton, G. N., Koshut, T. M., Mallozzi, R. S., Stollberg, M., & Lestrade, J. P. 1994, ApJS, 92, 229

Klebesadel, R. W. 1992, in Gamma-Ray Bursts. Observations, Analyses, and Theories, ed. C. Ho, R. I. Epstein, & E. E. Fenimore (Cambridge University Press:Cambridge), 161

Kouveliotou, C., Meegan, C. A., Fishman, G. J., Bhat, N. P., Briggs, M. S., Koshut, T. M., Paciesas, W. S., & Pendleton, G. N. 1993, ApJ, 413, L101

Mao, S., & Paczyński, B. 1992, ApJ, 388, L45

Mao, S., & Yi, I. 1994, ApJ, 424, L131

Norris, J. P., Nemiroff, R. J., Scargle, J. D., Kouveliotou, C., Fishman, G. J., Meegan, C. A., Paciesas, W. S., & Bonnell, J. T. 1994, ApJ, 424, 540

Piran, T. 1992, ApJ, 389, L45

Ulmer, A., & Wijers, R. A. M. J. 1994, ApJ, submitted

Wickramasinghe, W. A. D. T., Nemiroff, R. J., Norris, J. P., Kouveliotou, C., Fishman, G. J., Meegan, C. A., Wilson, R. B., & Paciesas, W. S. 1993, ApJ, 411, L55

Yi, I., & Mao, S. 1994, Phys. Rev. Lett., submitted

---

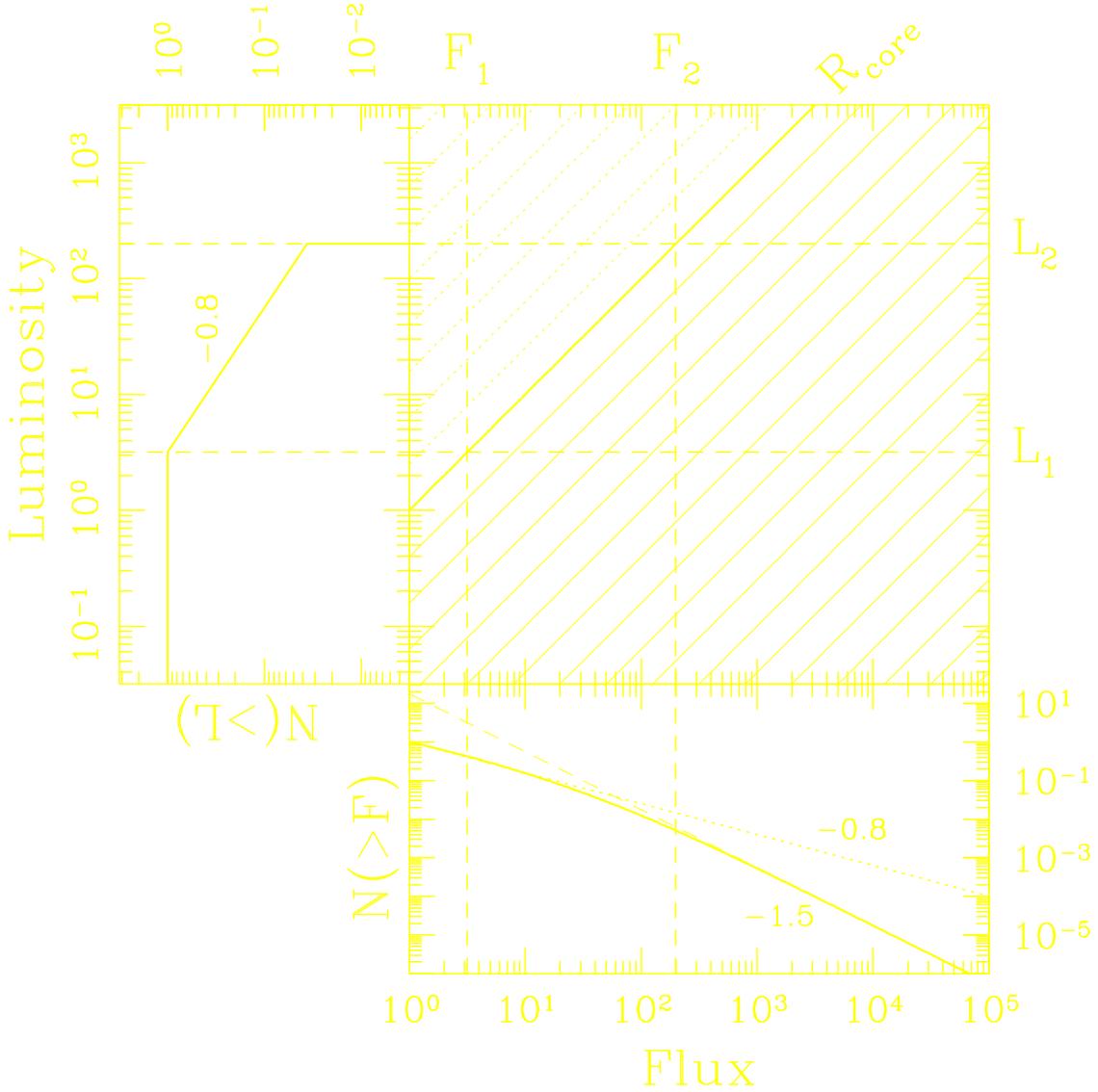

Fig. 1.— Illustration of how the observed $\log N(>F) - \log F$ arises from a convolution of density and luminosity distributions. Diagonal lines in the main panel represent shells of constant distance. For distances greater than $R_{\rm core}$ (dashed lines at top left) shells are progressively less populated, implying that samples of bursts with fluxes less than $F_2$ have relatively fewer intrinsically luminous bursts. This in turn means that the distribution of any quantity that is correlated with intrinsic luminosity differs in a sample of bursts that are brighter than $F_2$ from a sample of bursts that are fainter than $F_2$.

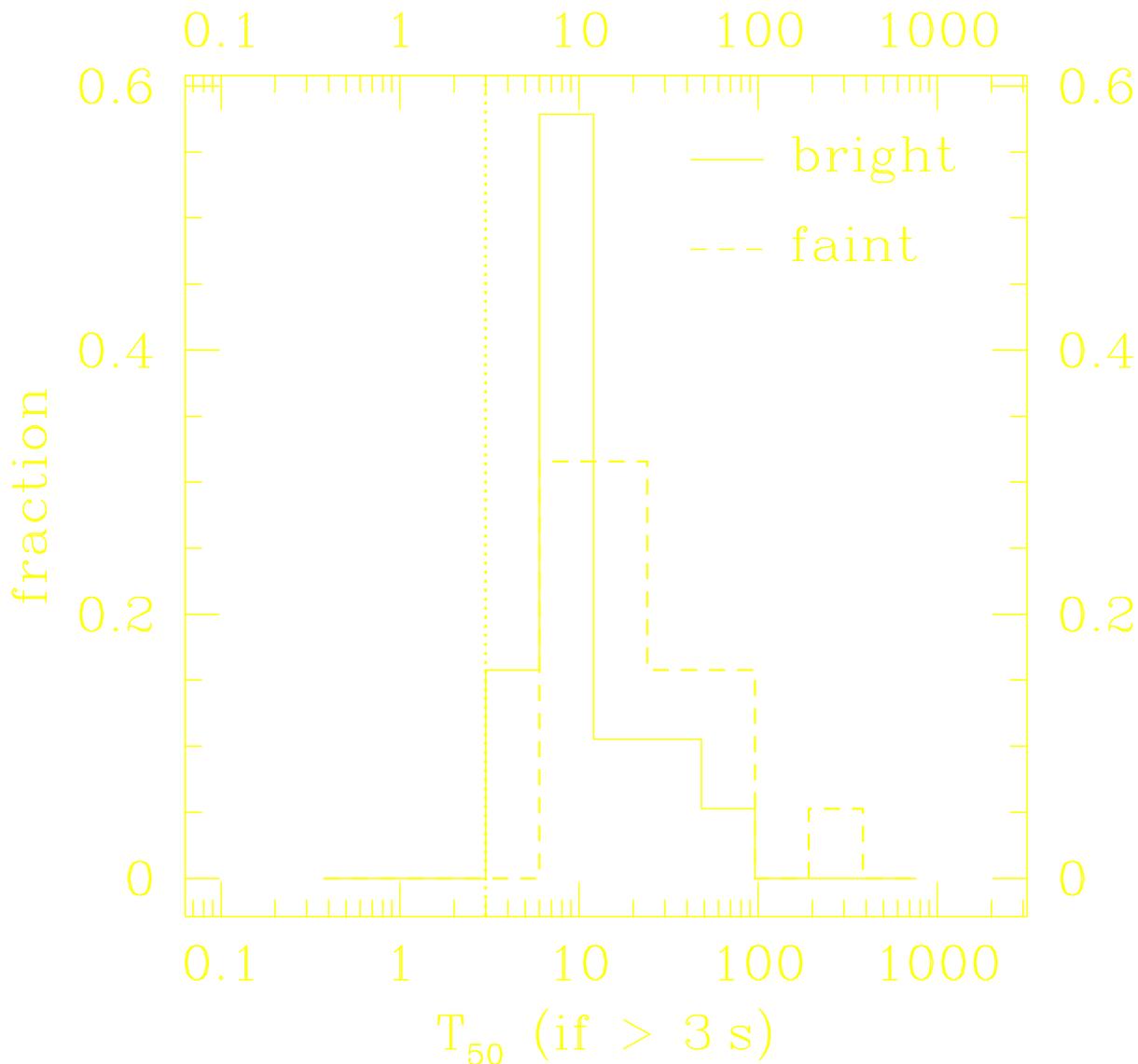

Fig. 2.— Duration comparison of a bright and faint sample of bursts from the second BATSE catalog. The brightness measure used is the peak count rate in 64 ms bins; the bright sample (solid line) consists of the 10% brightest bursts, and the faint sample (dashed line) consists of the bursts that rank between 10% and 20% from the bottom in a brightness-ordered list (each sample has 19 bursts). The bin width is $\log_{10}(2)$, so a duration change of a factor 2 means a one-bin shift between the two histograms. The dotted line marks the duration cutoff applied to both samples.